\documentclass[journal=jacsat,manuscript=article]{achemso}
\usepackage{chemformula} 
\usepackage[T1]{fontenc} 
\usepackage{graphicx}
\usepackage{dcolumn}
\usepackage{dcolumn}
\usepackage{bm}
\usepackage{color}
\usepackage{CJK}
\usepackage{fancyhdr}
\usepackage{calc}
\usepackage{float}
\usepackage{mpgmpar}
\usepackage{adjustbox}

\author{Junjun Jia}
\affiliation{Global Center for Science and Engineering (GCSE), Faculty of Science and Engineering, Waseda University, 3--4--1 Okubo, Shinjuku, Tokyo 169--8555, Japan.}
\email{jia@aoni.waseda.jp}

\author{Naoya Iwata}
\affiliation{Graduate School of Advanced Science and Engineering, Waseda University, 3–4–1 Okubo, Shinjuku, Tokyo 169--8555, Japan.}

\author{Masashi Suzuki}
\affiliation{Integrated Graduate School of Medicine, Engineering, and Agricultural Sciences, University of Yamanashi, Kofu 400-8511, Japan.}

\author{Takahiko Yanagitani}
\affiliation{Graduate School of Advanced Science and Engineering, Waseda University, 3–4–1 Okubo, Shinjuku, Tokyo 169--8555, Japan.}
\alsoaffiliation{Kagami Memorial Research Institute for Materials Science and Technology, Waseda University, 2–8–26 Nishiwaseda, Tokyo 169–0051, Japan.}
\alsoaffiliation{JST PRESTO, 4–1–8 Honcho, Kawaguchi, Saitama 332-0012, Japan.}
\email{yanagitani@waseda.jp}

\title[ACS Applied Materials & Interfaces]{Enhanced electromechanical coupling in Yb--substituted III--V nitride alloys}


\begin{document}

\begin{abstract}
Group-III nitride alloys are currently used in various microwave communication applications because of the giant enhancement in electromechanical coupling after alloying with rocksalt nitrides such as YbN or ScN. Herein, the Yb--substitution induced enhancement for electromechanical coupling in wurtzite III--V nitrides is studied via theoretical calculations and experiments. The substitution induced mechanical softening and local strain can enhance electromechanical coupling. The mechanical softening induced by Yb substitution shows less dependence on the parent AlN or GaN, which is considered to be caused by the Yb--Yb pair interaction in the $c$--axis, and the difference of electromechanical coupling between the GaN-- and AlN--based alloys mainly comes from the enhancement effect of Yb substitution for piezoelectric response. The largest change in piezoelectric response relative to the parent nitride is observed in GaN--based alloy, which is mainly considered as a consequence of small piezoelectric constant of the parent GaN. Our calculations also reveal that the substitutional element with a closer ionic size to the host cation is easier to substitute into the host nitride, and produces a larger internal strain to partly contribute to the enhancement in piezoelectric response. This can serve as a simple guideline to identify alloying components in a search for a massive increase in electromechanical coupling. 
\end{abstract}

\section{Introduction}

Group-III nitride material AlN is currently used widely in various microwave communication applications, such as GHz frequency filters in 5G applications, because of their high thermal stability, operating frequencies, $Q$ factor, and reliability.\cite{Yanagitani2014, Jia2021} For these applications, the small electromechanical coupling coefficient ($k_t^2$) of AlN can be enhanced by alloying with rocksalt nitrides such as ScN, YN, or YbN.\cite{Akiyama2009, Akiyama2009-2, Tholander2013, Tholander2015,Yanagitani2014} As an alternative III--V semiconductor, GaN shows a promising temperature coefficient of frequency (--12.6 ppm/$^{\circ}$C),\cite{Yanagitani2014, Jia2017} which is lower than that of AlN, and advantageous for application of frequency--stable resonators. However, GaN nitride alloy shows a relatively small $k_t^2$, about one--third of AlN nitride alloy,\cite{Yanagitani2014, Jia2021} which limits the resonator applications because the bandwidth and the filter insertion loss is determined by the $k_t^2$ value of the piezoelectric layer. It is therefore necessary to have a systematic theoretical and experimental investigation on the $k_t^2$ enhancement in GaN alloy.

Based on experimental observations, $k_t^2$ enhancement occurs in III--V nitride alloys only if the impurity element substitutionally occupies the cation site in the host film. For most AlN--based nitride alloys, such substitution often causes 1) an increase in field--induced strain $via$ an increase in the longitudinal piezoelectric constant $e_{33}$ and (2) a simultaneous decrease in the longitudinal elastic stiffness $c_{33}$.\cite{Tholander2013, Tholander2015, Jia2021} The same tendency is seen in GaN, but the $k_t^2$ in GaN--based nitride alloys is difficult to reach that of AlN alloys. Is such a difference an intrinsic feature for GaN or does it come from experimental fabrication conditions? So far, there is a lack of detailed explanation.  
 
Following the above enhancement mechanism, a large $k_t^2$ often requires high alloy concentrations, \textit{e.g.}, 42\% ScN or 11\% YbN in the AlN films,\cite{Akiyama2009, Jia2021} and thus a high concentration substitution, making it difficult to produce alloy films under equilibrium conditions. In practice, nonequilibrium physical vapor deposition techniques, such as sputtering, are widely employed to fabricate disordered solid solutions.\cite{Akiyama2009, Yanagitani2014, Mirko2016, Jia2014} Trumbore\cite{Trumbore1960} investigated the solid solubility of various impurities in silicon and germanium and proposed three factors that might affect the solubility: chemical compatibility, atomic size, and crystal structure. His solubility chart is still being accepted universally. For III--V parent nitrides, they have the same wurtzite structure, and the substitutional elements with valence state of 3+ are considered chemically compatible with their cation M$^{3+}$ (Al$^{3+}$, Ga$^{3+}$ or In$^{3+}$). Generally speaking, the substitutional element with a similar ionic size to the host cation is considered easy to substitute into the host cation site. Therefore, using the same substitutional element is a good way to investigate the size effect on the enhancement of electromechanical coupling in different III--V parent nitrides.  

Ga$^{3+}$ has a larger ionic size (47.0 pm) than Al$^{3+}$ (39.0 pm) in a tetrahedron; both are smaller than those of well--known substitutional elements, such as Cr$^{3+}$ (61.5 pm), Sc$^{3+}$ (74.5 pm), Yb$^{3+}$ (86.8 pm), and Y$^{3+}$ (90.0 pm), where ionic sizes with 6 coordination number are selected for the substitutional elements.\cite{Shannon1976} Based on these ionic radii, relatively speaking, Yb is considered to easily substitute into GaN. Our sputtering experiments have demonstrated this point that 1) Yb alloying concentration is much higher in GaN than one in AlN;\cite{Yanagitani2014} 2) $k_t^2$ disappears above 25\% YbN in the sputtered (Yb,Al)N films, and such disappearance occurs above 32\% YbN in the sputtered (Yb,Ga)N films.\cite{Yanagitani2014} Surprisingly, although GaN alloy can reach higher alloy concentration, (Yb,Ga)N alloy only has a maximal $k_t^2$ value of 3.2\%,\cite{Yanagitani2014} which is obviously lower than that of (Yb,Al)N.\cite{Jia2021} The difference in $k_t^2$ between them may come from the difference between $d$ and $s$ orbits, becasue Al has a [Ne]3$s^2$3$p^1$ configuration, different from [Ar]4$s^2$3$d^{10}$4$p^1$ of Ga from the perspective of electronic structure. In this paper, we will show that although the calculations with the semicore $d$ electrons reproduces the expeiments well, such treatment cannot explain the difference in $k_t^2$ between Yb--substituted AlN and GaN alloys. Thus, a question arises as to how the local stiffness or piezoelectricity is to be enhanced by introducing the YbN endmember into GaN? 

To the best of our knowledge, there is no study on the interplay between substitutional element and parent wurtzite structure, and it remains yet unknown which of these two effects will dominate the electromechanical coupling. In this study, we focus on the influence of Yb substitution on electromechanical coupling in different parent wurtzite nitrides, including InN, by employing the experiments and density--functional theory (DFT). First principles calculations for the elastic constant, piezoelectric constants, and dielectric constant are also presented.

\section{Experiments and Calculations} 

(Yb, Ga)N films with the $c$--axis orientation were fabricated on Ti electrode film/silica glass substrate by sputtering deposition, where a highly oriented (001) Ti electrode film was used as the under--layer electrode to form a HBAR structure. The film structure was characterized by X--ray diffraction (XRD) with a $2\theta$--$\omega$ configuration (X'Pert PRO, PANalytical). The Yb concentrations were analyzed by electron probe microanalysis (JXA-8230, JEOL). The $k_t^2$ values were determined by comparing the experimental and theoretical longitudinal wave conversion loss curve versus the frequency of HBAR resonators.\cite{Jia2021,Yanagitani2014}  

Theoretical calculations for (Yb,Ga)N and (Yb, In)N  have been extensively described in our previous study.\cite{Jia2021} The Vienna \textit{ab initio} simulation package (VASP) is used for structural optimization with the generalized gradient approximation as parameterized by Perdew et al. for the exchange--correlation potential.\cite{Kresse1996, PBE} A 3$\times$3$\times$2 supercell is selected, and the special quasirandom structure (SQS) method is employed to simulate the random distribution of Yb in the parent wurtzite structure with different Yb concentrations. The SQSs are generated by optimizing the locations of Yb atoms to minimize the Warren--Cowley pair short--range order parameters,\cite{Zunger1990} which are calculated up to the sixth coordination shell. In order to realistically simulate the chemical disorder of actual nitride alloys, we constructed approximately 10 SQS structures with the same Yb concentration, and selected the structure with the lowest total formation energy to calculate piezoelectric and elastic tensors of (Yb,Ga)N and (Yb,In)N. The wurtzite (Yb,Ga)N and  (Yb,In)N structures are first relaxed using DFT. The Monkhorst--Pack $k$--point grids are set to 3$\times$3$\times$3. The plane--wave basis set with a cutoff energy of 600 eV is adopted, and the total energies converge to less than 10$^{-9}$ eV for structural relaxation. The elastic and piezoelectric tensors are calculated using density--functional perturbation theory.

\section{Results and Discussions} 

\subsection{Structural properties} 

\begin{figure}[h]
\begin{center}
\includegraphics[clip, width=11.0cm]{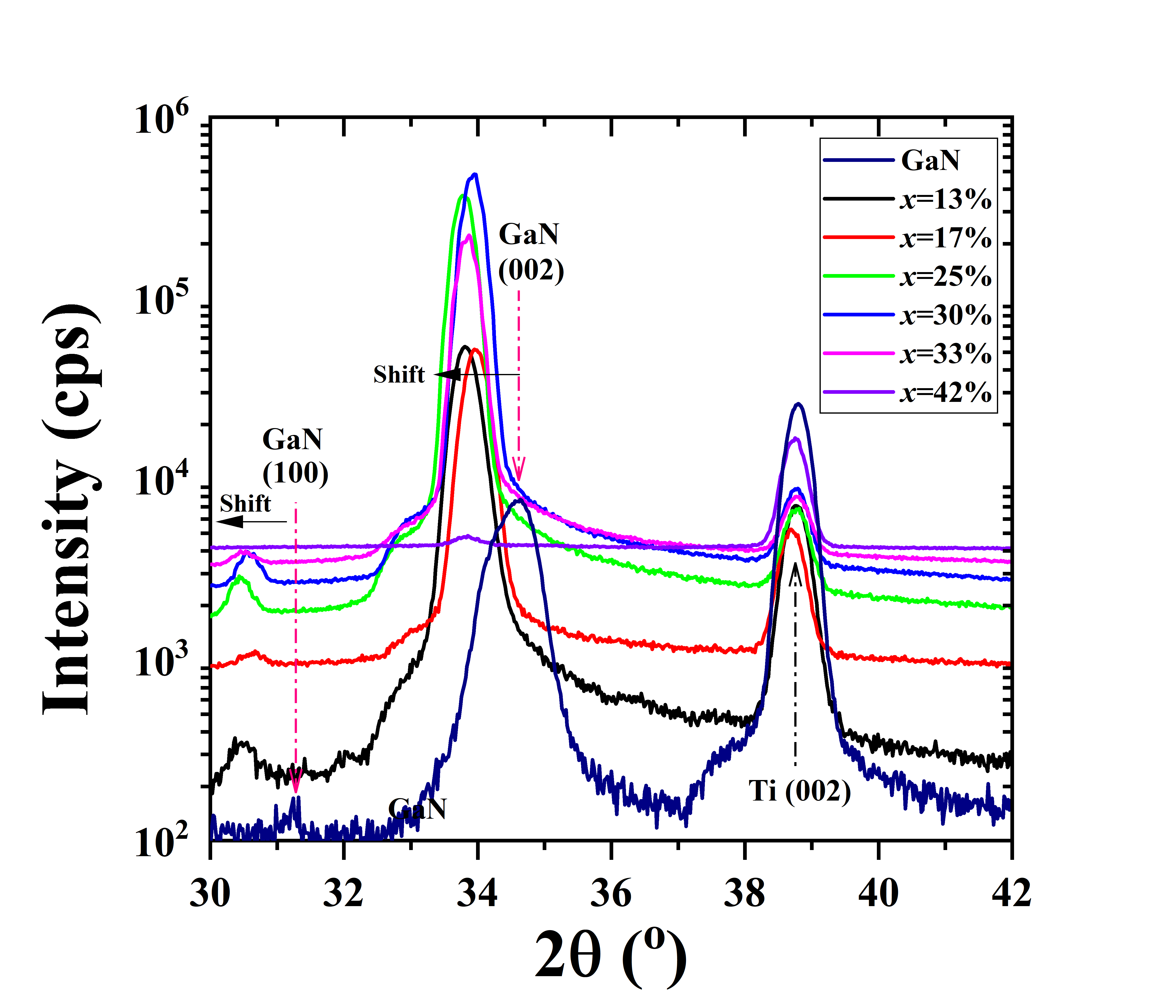}
\caption{Diffraction peaks of (Yb,Ga)N films with different Yb concentrations, which were deposited on the Ti electode film/glass substrate for HBAR applications. Herein, $x$=Yb/(Yb+Ga) is the concentrations of Yb in the formed binary nitride alloy.}
\label{xrd} 
\end{center}
\end{figure}

Figure \ref{xrd} shows the 2$\theta/\omega$ XRD patterns of the (Yb,Ga)N films with different Yb concentration (Yb$_x$Ga$_{1-x}$N). The observed peaks at 31.2$^\circ$, 34.6$^\circ$, and 38.8$^\circ$ are diffractions from the GaN (100), GaN (002), and Ti (002) planes, respectively.\cite{Jia2017} At $x\leq$33\%, all the films show high $c$--orientation along the thickness direction on the Ti electrode. Due to the presence of GaN (100) peaks in the 2$\theta/\omega$ XRD patterns, the deposited (Yb,Ga)N films are ploycrystalline. Whereas, a drastic decrease in the peak intensity of GaN (002) is observed at $x$>33\%, suggesting the degradation of wurtzite structure with alloying more YbN. In addition, the GaN (100) and (002) peaks show obvious shift toward the lower angle side with the Yb concentration, indicating the increase in both $a$-- and $c$--lattice spacings. 

Before starting piezoelectric calculations, it is crucial to perform calculations for its bulk phase to determine the structural properties for (Yb,Ga)N films. We performed structure optimization calculations of wurtzite GaN crystal with LDA (using only 4$s$ and 4$p$ valence orbitals), PBE (using only 4$s$ and 4$p$ valence orbitals), GGA+$d$ (including 4$d$ orbitals). Table \ref{GaN-latt} lists the lattice spacings in the $a$-- and $c$--axis. Our optimized lattice parameters of wurtzite GaN are $a$=3.247 (3.187) \AA\ and $c$=5.281 (5.185) \AA\ in PBE (LDA), the calculated values using GGA+d fall in--between the LDA and PBE data. We find that the LDA lattice parameters are very close to experimental values. This is very common in the DFT calculations. Meanwhile, the lattice parameters with PBE and GGA+$d$ give 1.8\% and $\sim$0.8\% larger value than experimental ones. 

\begin{table*}[h]
\caption{Calculated structural parameters of (Yb, Ga)N based on first principles calculations using different functionals. For comparison, values obtained from other theoretical studies and experiments are also included.}
\begin{tabular}{@{}llllll}
\hline
    & $a$ (\AA)  & $c$ (\AA) & $c/a$   & $\rho$ (g/cm$^3$) &   \\
\hline
AlN\cite{Jia2021}  & 3.128     & 5.015  & 1.603  & 3.20 & PBE  \\
GaN  & 3.189\cite{Edgar1994}      & 5.185\cite{Edgar1994}  & 1.624  & 6.07\cite{Nakamura2012} & Exp. \\ 
GaN  & 3.187     & 5.185 & 1.627  & 6.10 & LDA  \\ 
GaN        & 3.217     & 5.243 & 1.630  & 5.92 & GGA+$d$  \\ 
GaN        & 3.247     & 5.281 & 1.626  & 5.77 & PBE  \\ 
\hline
\end{tabular} 
\label{GaN-latt}
\end{table*}

\begin{figure}[h]
\begin{center}
\includegraphics[clip, width=12.0cm]{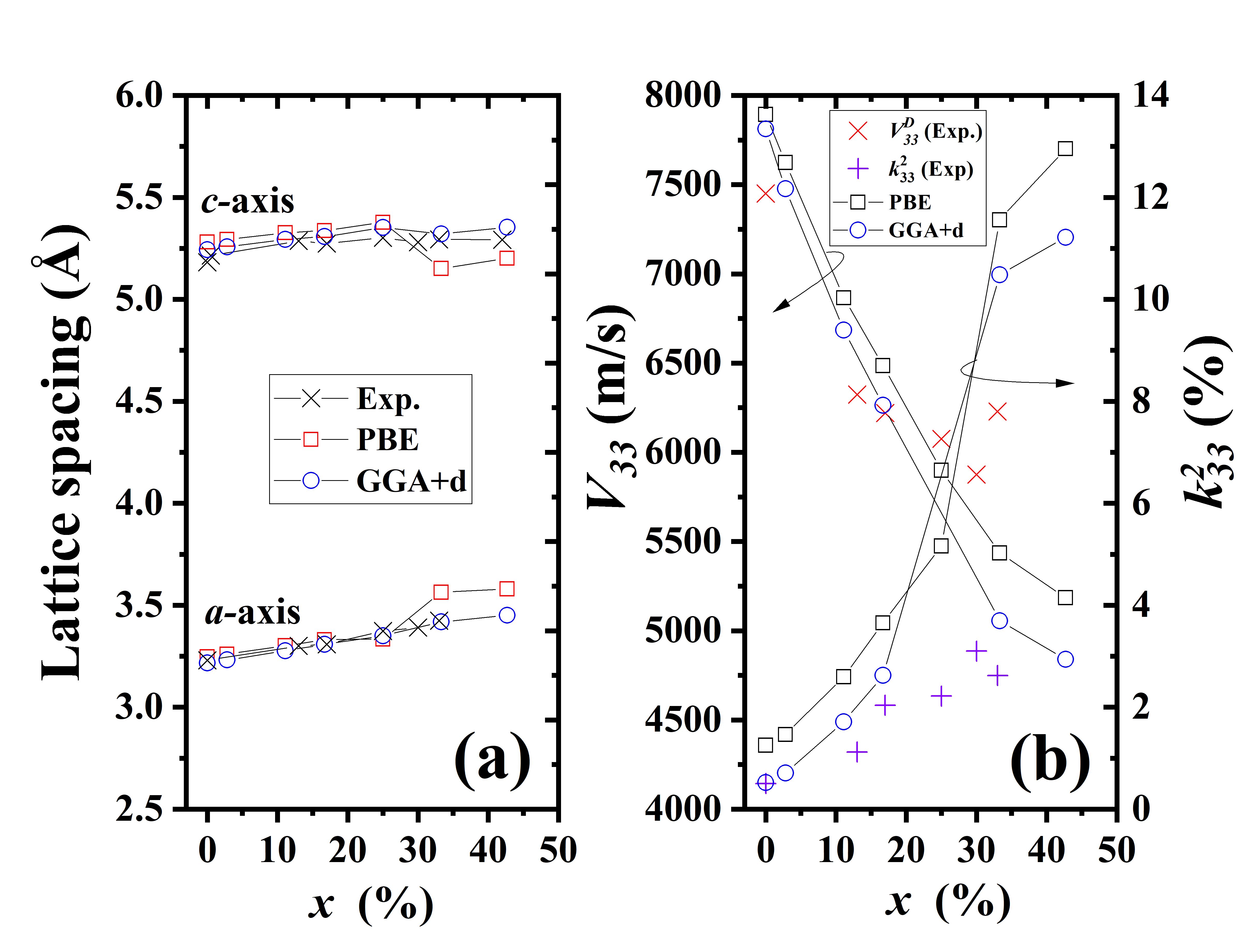}
\caption{The measured lattice spacings in the $a$-- and $c$--axis (a) and measured $V_{33}^D$ and $k_{33}^2$ (b) of sputtered (Yb,Ga)N films are compared with those calculated by first principles calculations with PBE and GGA+$d$, respectively. The cross and plus symbols represent the measurement values, and the open circles or squares denote the calculated values. $V_{33}^D$ is the shear acoustic wave velocities.}
\label{exp}
\end{center}
\end{figure}

Fig. \ref{exp} shows the calculated lattice constants of (Yb,Ga)N. The $a$--axis lattice spacing monotonically increases with increasing Yb concentration, whereas the $c$--axis lattice spacing initially increases and subsequently decreases with increasing Yb concentration. As shown in Fig. \ref{xrd}, XRD patterns show a similar tendency, where the (0002) plane gradually shifts toward the low angle side. The extrapolated $a$-- and $c$--lattice constants from the XRD patterns agree with the results from our first principles calculations with GGA+$d$ potential, which tends to be slightly overestimated but systematically. We prefer to display the results only given by GGA+$d$ potential in this study due to the reliable values given by the functional.

\subsection{Phase stability and local structure of Yb}

\begin{figure}[h]
\begin{center}
\includegraphics[clip, width=12.0cm]{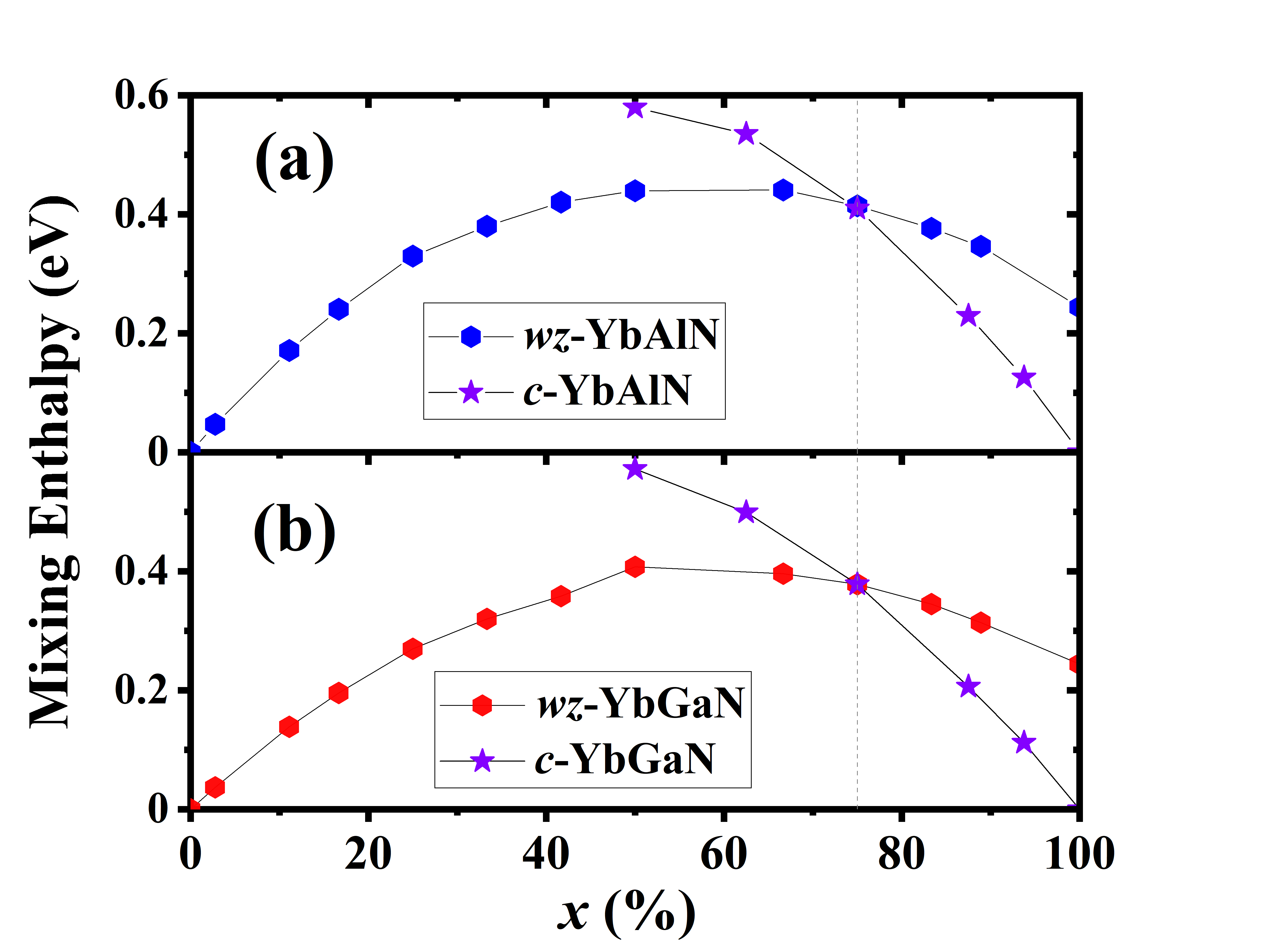}
\caption{Mixing enthalpies $\Delta H_{\mathrm{mix}}$ of (Yb,Al)N (a) and (Yb,Ga)N (b) alloys in the hexagonal and rocksalt phases. The dashed lines show the wurtzite--to--rocksalt phase transition points.}
\label{Entalpy}
\end{center}
\end{figure}

Toward practical applications for SAW or FBAR devices, Yb must be substituted into the cation site of the wurtzite structure to maintain the piezoelectricity. The maximal substitution concentration corresponds to the solid solubility.\cite{Trumbore1960} However, upon excess substitution, the wurtzite structure may become unstable due to the interaction between Yb substitutional atoms. To understand the phase stability, we calculated the thermodynamic free energy of Yb-substituted nitride alloys. The mixing Gibbs free energy $\Delta G_{\mathrm{mix}}$ was calculated using 72--atom SQS supercells for each composition as follows:\cite{Alling2007, Talley2018} 

\begin{equation}
\begin{split}
\Delta G_{\mathrm{mix}} (\mathrm{Yb}_x \mathrm{M}_{1-x}\mathrm{N})=\Delta H_{\mathrm{mix}} (\mathrm{Yb}_x \mathrm{M}_{1-x}\mathrm{N}) \\- T\Delta S(\mathrm{Yb}_x \mathrm{M}_{1-x}\mathrm{N}),
\label{gibbs}
\end{split}
\end{equation}

where $x$ is the mole fraction of YbN,  $T$ is the temperature, and M denotes the group--III metal Al or Ga. Besides, $\Delta H_{\mathrm{mix}}$ is the mixing enthalpy and is expressed as follows:\cite{Alling2007, Zhang2013}

\begin{equation}
\begin{split}
  \Delta H_{\mathrm{mix}} (\mathrm{Yb}_x \mathrm{M}_{1-x}\mathrm{N}) = E(\mathrm{Yb}_x \mathrm{M}_{1-x}\mathrm{N}) \\- xE(\mathrm{YbN}, c)
  -(1-x)E(\mathrm{MN}, wz), 
\label{Hmix}
\end{split}
\end{equation}

where $E$ is the total energy per formula unit. $\Delta S$ corresponds to the configuration entropy of an ideal solution (the mean--field approximation) and is expressed as follows:

 \begin{equation}
  \Delta S (\mathrm{Yb}_x \mathrm{M}_{1-x}\mathrm{N}) = -k_B \left [ x\mathrm{ln}x + (1-x)\mathrm{ln}(1-x) \right], 
\label{entropy}
\end{equation}

where $k_B$ is the Boltzmann constant. Fig. \ref{Entalpy} shows the calculated $\Delta H_{\mathrm{mix}}$ for the wurtzite and rocksalt phases of (Yb,Al)N and (Yb,Ga)N. Wurtzite structures are favored up to $x$$\sim$0.75 in both (Yb,Al)N and (Yb,Ga)N, whereas rocksalt structures are favored at higher Yb concentrations. The wurtzite--to--rocksalt phase transition point for (Yb,Al)N is the same as that of (Yb,Ga)N. This seems inconsistent with our empirical criterion, where because the ionic size of Ga$^{3+}$ is closer to that of Yb$^{3+}$, the solubility of Yb in wurtzite GaN is generally expected to be slightly larger than that in wurtzite AlN. Further analysis shows that $\Delta H_{\mathrm{mix}}(\mathrm{Yb}_x \mathrm{M}_{1-x}\mathrm{N})>\Delta H_{\mathrm{mix}}(\mathrm{MN})$ (Fig. \ref{Entalpy}), suggesting that alloying YbN into wurtzite AlN or GaN is an endothermic reaction. In Fig. \ref{Entalpy}, $\Delta H_{\mathrm{mix}}$ in wurtzite (Yb, Ga)N is smaller in energy than that of the wurtzite  (Yb,Al)N, indicating that (Yb,Ga)N is easy to form. It is worthy to note that the wurtzite--to--rocksalt phase transition points for (Yb,Al)N and (Yb,Ga)N are obviously larger than our experimental observations.\cite{Jia2021}  Our experiments indicate that the wurtzite phase is found to be favourable up to $x$=0.25 and $x$=0.32 for sputtered (Yb,Al)N and (Yb,Ga)N films, respectively, beyond which electromechanical coupling disappears in the measurements. Such a difference between the calculated phase transition points and experimental observations is considered from the phase instability of nitride alloys, and further confirmation from the calculation of Gibbs free energy is necessary. 

\begin{figure}[h]
\begin{center}
\includegraphics[clip, width=12.0cm]{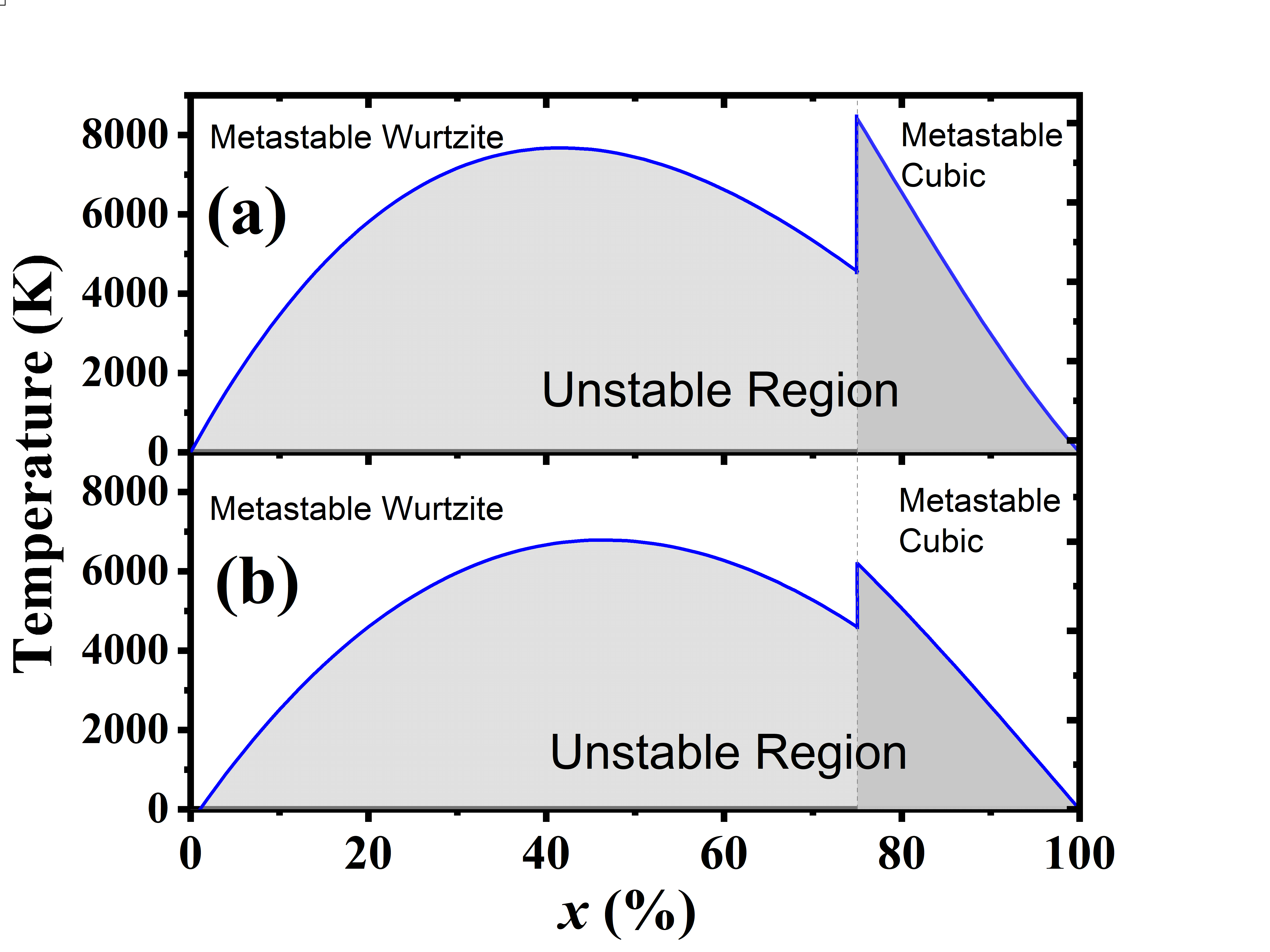}
\caption{Calculated temperature--composition phase diagram for (Yb,Al)N (a) and (Yb,Ga)N (b) alloys in hexagonal and rocksalt phases. The filled patterns represent unstable spinodal decomposition regions. The dashed lines show the wurtzite--to--rocksalt phase transition points.}
\label{Phase}
\end{center}
\end{figure}

The information about the stability of nitride alloy can be obtained from the Gibbs free energy versus composition diagram by a common tangent construction. Our calculations show $\Delta G_{\mathrm{mix}}(\mathrm{Yb}_x \mathrm{M}_{1-x}\mathrm{N}, wz)>\Delta G_{\mathrm{mix}}(\mathrm{MN}, wz)$, suggesting that nitride alloys are thermodynamically unstable, and the spinodal decomposition spontaneously occurs at $\frac{d^2\Delta G_{\mathrm{mix}}}{dx^2}<0$. Because $\Delta H_{\mathrm{mix}}$ and $\Delta S$ are calculable, a phase diagram of the spinodal decomposition temperature and composition can be constructed to study phase stability using Eq. \ref{gibbs}, as proposed by Talley {\it et al.}.\cite{Talley2018}. Fig. \ref{Phase} shows that most alloy compositions lie within the spinodal region at the typical film deposition temperature, \textit{e.g.}, 300--900 K, which is considered to spontaneously decompose toward their endmembers.\cite{Zukauskaite2012} Such unstable alloy films require nonequilibrium conditions to fabricate as the sputtering deposition does. This is because the supercooling of sputtered atoms on the substrate and high deposition rate could kinetically limit the diffusion--controlled spinodal decomposition for high Yb concentrations. Moreover, the similar temperature--composition phase diagram is also obtained for (Al,Sc)N, indicating that the (Al,Sc)N alloy films grown at the typical deposition temperature are located in thermodynamic conditions deep within the spinodal region.\cite{Talley2018} Talley {\it et al.} suggests that such thermodynamical unstability is associated with the decomposition of (Al,Sc)N films in the postannealing process.\cite{Talley2018, Hoglund2009}. Furthermore, Fig. \ref{Phase} also shows that the spinodal temperature of (Yb,Ga)N is lower than that of (Yb,Al)N, suggesting that (Yb,Ga)N is more stable at the same substitutional concentration. 

Notably, $\Delta G_{\mathrm{mix}}$ in Eq. (\ref{gibbs}) is an approximation for the change in real Gibbs free energy. This is because 1) $\Delta H_{\mathrm{mix}}$ is obtained from first principles calculations at 0 K and its internal energy term does not include the temperature effect via kinetic energy of atomic vibration, and 2) the second term on the right--hand side of Eq. (\ref{gibbs}) neglects the vibrational entropy. However, in Fig. \ref{exp} (b), the measured longitudinal acoustic wave velocity $V_{33}^D$ at room temperature almost agrees with the calculated $V_{33}$ at 0 K, which implies that the acoustic phonon has less temperature dependence, allowing us to expect that the temperature--induced change in internal energy is sufficiently small up to room temperature. Moreover, the vibrational entropy is considered less relevant than configurational entropy because the difference between the measured and calculated lattice spacings is fairly small as shown in Fig. \ref{exp} (a).   

The calculations of mixing Gibbs free energy provide insights into phase stability of wurtzite alloy structures to maintain the piezoelectricity. Moreover, from the microscopic perspective, the interatomic interaction between substitutional atoms becomes crucial for heavy substitution and affects the spatial configuration of Yb substitution in the parent wurtzite AlN or GaN. For any two substitutional Yb atoms in the wurtzite structure, the interatomic interaction between Yb atoms can be evaluated by calculating their pair interaction energies as follows:\cite{Jia2021} 

\begin{equation}
\begin{split}
  \delta E^{(n)}=\left[E(\mathrm{M}_{(\frac{m}{2}-2)}\mathrm{Yb}_2\mathrm{N}_{\frac{m}{2}}) + E(\mathrm{M}_{\frac{m}{2}}\mathrm{N}_{\frac{m}{2}})  \right] \\
-2E(\mathrm{M}_{(\frac{m}{2}-1)}\mathrm{Yb}\mathrm{N}_{\frac{m}{2}}),
\label{pair}
\end{split}
\end{equation}

where M=Al, Ga, and In, $n$ denotes the pair index, and $m$=72 is the number of atoms in the supercell. The above formula describes the change of internal energy when two isolated substitional Yb atoms are brought close as the $n$th nearest neighbor from an infinite separation. The total energies $E$ correspond to the fully relaxed supercell calculations. In the calculations, an identical $k$--mesh (3 $\times$ 3 $\times$ 3) is applied for all three total energies for consistency.

\begin{figure}[h]
\begin{center}
\includegraphics[clip, width=12.0cm]{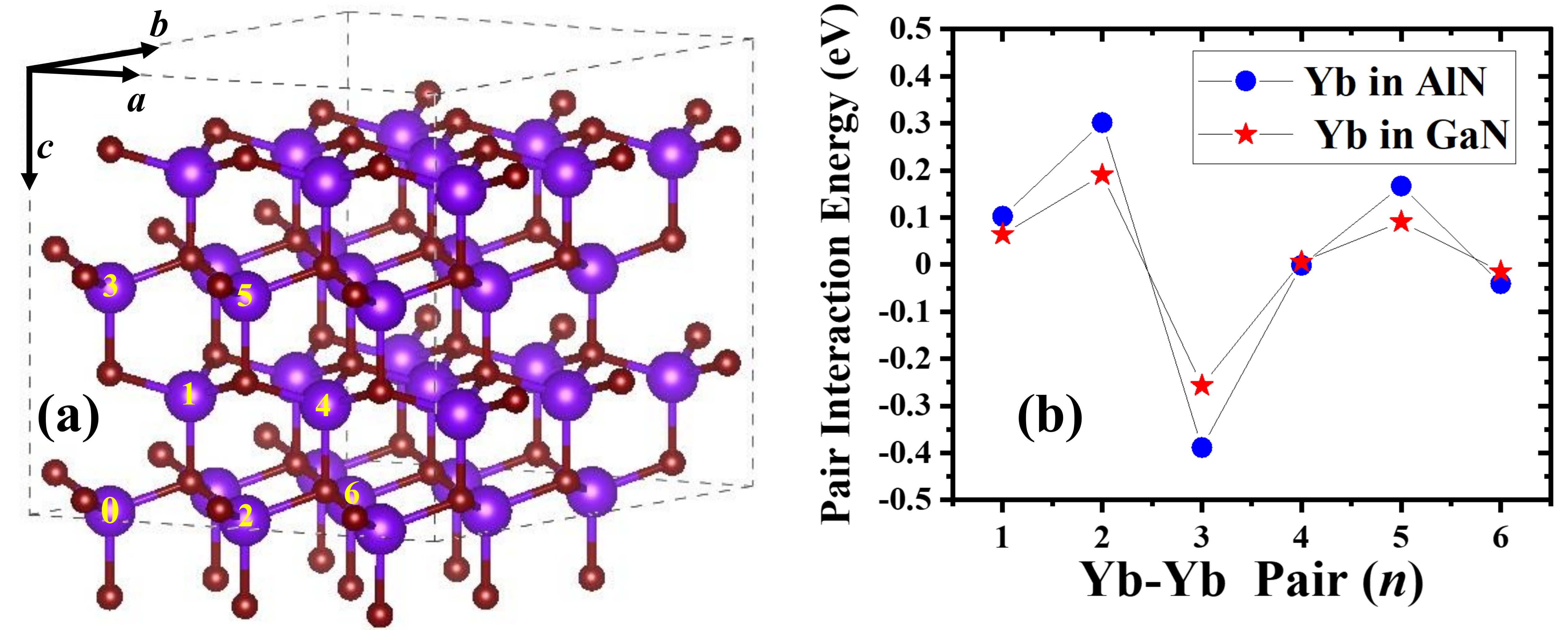}
\caption{(a) The used pair configurations in a 3$\times$3$\times$2 supercell with $m$=72 atoms (purple: cations; brown: N ions). (b) The pair interaction energies $\delta$$E$($n$) between the two $n$th nearest neighbor Yb atoms, where $n$ = 1, 2, 3, 4, 5, 6 neighbor. The pair interaction energy $E(\mathrm{M}_{(\frac{m}{2}-2)}\mathrm{Yb}_2\mathrm{N}_{\frac{m}{2}})$ is calculated by placing a Yb atom at the origin and the other at the $n$ position. For comparison, $\delta$$E$($n$) values in (Yb,Al)N are also cited.\cite{Jia2021}}
\label{pairinteraction}
\end{center}
\end{figure}

In Fig. \ref{pairinteraction}, our results suggest that Yb--Yb pairing avoids $n$=1, 2, 5 neighbor shells, and favors $n$=3, 4, 6 neighbor configurations, especially $n$=3. The negative pair energy at $n$=3 means the attractive interaction between the Yb--Yb pair, implying that it is easy to form --Yb--N--Yb-- bonds along the $c$ axis, which is expected to cause the elastic softening along the $c$ axis. The similar trend is also observed in (Yb,Al)N and (Sc,Al)N.\cite{Jia2021}

\subsection{Electromechanical Coupling}
So far, we have discussed the solubility of Yb in wurtzite AlN or GaN structure. Yb substitution is also expected to cause the local distortion in the wurtzite structure due to its size difference from the host cation. A question arises as to whether the distortion can be enhanced or redeemed for electromechanical coupling. To answer this question, let us start from the definition of the electromechanical coupling coefficient of piezoelectric material. It relates to the conversion rate between electrical energy and mechanical energy and is regarded as a figure of merit commonly used in piezoelectric device design. In our experiment, $k_t^2$ was extracted from the minimum value of HBAR conversion loss. The measured $k_{t}^2$ is known to be approximately equal to the longitudinal electromechanical coupling coefficient $k_{33}^2$ for $c$-oriented III--V semiconductor films:\cite{Feneberg2007, Jia2021} 

\begin{equation}
k_{t}^2 \sim k_{33}^2 = \frac{e_{33}^2}{\epsilon_{33} c_{33}},
\label{k33}
\end{equation}

where $e_{33}$ is the longitudinal piezoelectric constant, $\epsilon_{33}$ is the dielectric constant under a strain--free condition, and $c_{33}$ is the longitudinal elastic stiffness. In Fig. \ref{exp} (b), the measured $k^2_t$ values are compared with the calculated values. The $k_{33}^2$ values calculated with GGA+$d$ are good agreement with the experimentally measured $k_{t}^2$ up to approximately $x$=0.20. At x$>$0.20, $k_{33}^2$ becomes larger than $k^2_t$. One reasonable explanation is that alloying more YbN into GaN can affect the crystallinity, crystal orientation, and polarization direction of those sputtered films with the wurtzite structure as shown in Fig. \ref{xrd}, which lead to a smaller $k^2_t$ than that in the idealized structure. In addition, the repulsive Yb--Yb pair interactions on the basal plane is also considered to hinder further Yb substitution and thus result in a low $k^2_t$, as revealed by the calculations of the pair interaction energies in Fig. \ref{pairinteraction}. Based on Eq. (\ref{k33}), the Yb substitution into the wurtzite structure is expected to change the piezoelectric constant, elastic stiffness, and dielectric constant. These parameters can be evaluated by first principles calculations. Note that the $k_{33}^2$ values calculated with PBE functionals also reproduce the experimental tendancy well, suggesting that the existence of Gd $d$ electrons is not a dominant factor to affect $k_{33}^2$ in GaN alloys.

\begin{table}[ht]
\caption{\label{cij}Elastic ($c_{ij}$) and piezoelectric ($e_{ij}$) constants of (Yb,Ga)N at different Yb concentrations $x$, where $c^{E}_{33}$ is the experimental values by measuring (Yb,Ga)N films.\cite{Yanagitani2014} Elastic constants are in GPa, and piezoelectric constants are in C/m$^2$. The $c_{ij}$ and $e_{ij}$ values for GaN calculated with PBE from other groups are provided for comparison, and our calculations agree well with previous ones.\cite{Bernardini1997} Notably, $\epsilon_{33}$ is the 33 component of the dielectric tensor.}
\begin{adjustbox}{width=1\textwidth}
\begin{tabular}{lrlllllcccc}
\hline
  & $\epsilon_{33}$ & $c_{11}$ & $c_{12}$ & $c_{13}$ &  $c_{33}$  & $c_{44}$ & $c_{66}$ & $e_{31}$ & $e_{33}$ & $k^2_{33} (\%)$ \\
\hline \\
GaN            & 11.57  & 320  & 109 & 79   & 361 &  88    & 105      & -0.27  & 0.44 & 0.52\\
$x$=0.03      & 11.71  & 310  & 110 & 82   & 338 &  86    & 100       & -0.28  & 0.50 & 0.71\\
$x$=0.11      & 12.24  & 275  & 109 & 93   & 287 &  76    & 85         & -0.33  & 0.73 & 1.71\\
$x$=0.17      & 12.48  & 258  & 109 & 95   & 261 &  71    & 74         & -0.32  & 0.87 & 2.62\\
$x$=0.33      & 14.02  & 227  & 120 & 104   & 187 &  67    & 65        & -0.44  & 1.56 & 10.48\\
$x$=0.42      & 14.07  & 209  & 115 & 107   & 178 &  50    & 63        & -0.47  & 1.58 & 11.21\\
References for GaN (PBE) \\
GaN (PBE in this study)            & 11.04  & 328  & 128 & 95   & 355 &  86 & 100        & -0.35  & 0.66 & 1.25\\
{\it Calculated by Zhang et al.}\footnotemark[1]  & --  & 344  & 134 & 102 & 372 & 90 & 105   & -- & --& --\\
{\it Calculated by Bernardini et al.}\footnotemark[2]  & --  & -- & -- & -- & -- & -- & -- &-0.37  & 0.67 & --\\
\hline \\
\end{tabular}
\end{adjustbox}
\footnotetext[1]{Reference.~\cite{Zhang2013}.}
\footnotetext[2]{Reference.~\cite{Bernardini1997}.}
\end{table}

Table \ref{cij} shows that the calculated $c_{33}$ decreases rapidly while the calculated $\epsilon_{33}$ and $e_{33}$ increase with increasing YbN content in GaN. The obvious changes in $e_{33}$ and $c_{33}$ are mainly considered to cooperate to enhance the electromechanical coupling in the wurtzite GaN. Moreover, for (Yb,Ga)N,  elastic constants $c_{11}$, $c_{33}$, $c_{44}$ and $c_{66}$ were found to decrease while $c_{13}$ increased slightly with increasing YbN content. Furthermore, we calculated the longitudinal wave velocity $V_{33}$ via $V_{33}=\sqrt{c_{33}/\rho}$ with the density $\rho$, and then compared them with the experimentally determined $V_{33}^D$ values. Fig. \ref{exp} (b) shows that the calculated $V_{33}$ values at 0 K are consistent with the measured $V_{33}^D$ values at room temperature up to $x$=0.2. Such quantitative agreement allows us to expect that the acoustic phonon contribution to piezoelectricity in (Yb,Al)N alloys is small up to room temperature.\cite{Jia2021}

Next, let us discuss the change in $c_{33}$ and $e_{33}$ induced by Yb substitution. The change in $c_{33}$ due to the Yb substitution in AlN is partially attributed to the interaction between the Yb--Yb pair interaction along the $c$--axis in our previous study.\cite{Jia2021} In Fig. \ref{pairinteraction}, the similar behavior is observed in (Yb,Ga)N along the $c$--axis, implying that the mechanical softening is still partially driven by the Yb--Yb pair interaction, which originates from the formation of --Yb--N--Yb-- bonds in the $c$ axis. On the other hand, the substitution with impurity ions is expected to generate an internal strain in the host wurtzite structure, which induces the piezoelectric polarization and thus changes $e_{33}$. The macroscopic piezoelectric tensor coefficient along the $c$ axis has been formulated as follows:\cite{Bernardini1997, Wang2021} 

\begin{equation}
\begin{aligned}
e_{33} (x) = e_{33}^{\mathrm{clamped}} (x) + e_{33}^{\mathrm{int}} (x)  \\
= \frac{\partial P_3}{\partial\sigma} \bigg|_u +  \frac{4eZ_{33}(x)}{\sqrt{3}a(x)^2}\frac{\partial u}{\partial \sigma},
\end{aligned}
\label{eqn-e33}
\end{equation}

where the first term denotes an electronic response to the external strain ($clamped$--$ion$ term), which is evaluated by freezing the atomic coordinates at their zero--strain equilibrium positions, and the second term quantifies the effects of internal strain on the polarization ($e_{33}^{\mathrm{int}}$), arising from the distortion of the ionic coordinates at fixed strain. In the second term, $e$ is the elementary charge, $a$ represents the equilibrium lattice parameter, $Z_{33}$ represents the dynamical Born charge in units of $e$, $\sigma$ is the macroscopic applied strain along the $c$ direction, and $u$ is the wurtzite internal parameter. Note that the wurtzite structure comprises two interpenetrating hexagonal sublattices for cation and anion, respectively, where each sublattice contains four atoms per unit cell. These two sublattices are displaced with respect to each other along the $c$--axis by the amount $u$, known as the internal parameter ($u$=0.5 for the hexagonal phase and $u\neq$0.5 for the wurtzite phase). 

\begin{figure}[h]
\begin{center}
\includegraphics[clip, width=12cm]{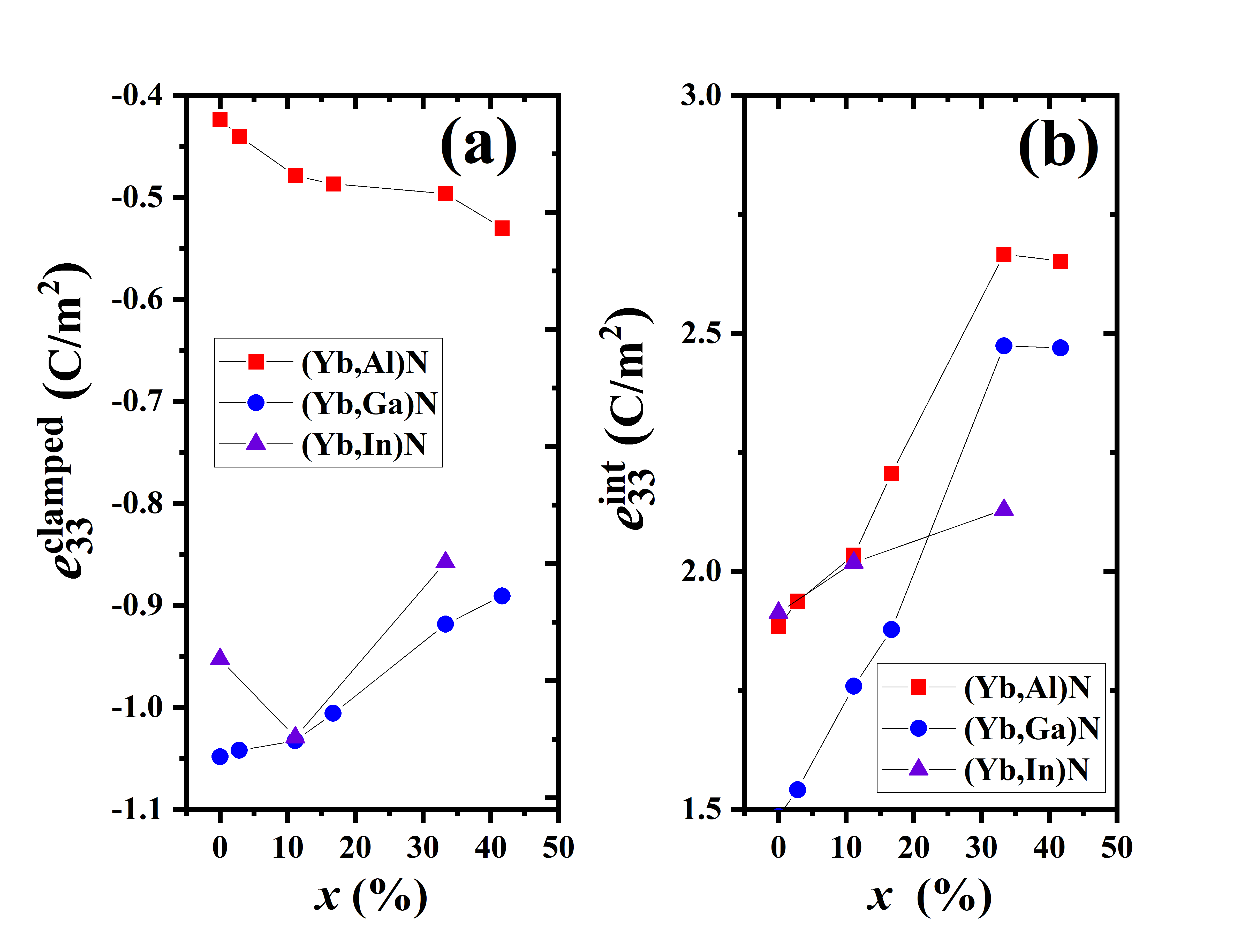}
\caption{Calculated contributions of $e_{33}$ introduced in Eq. \ref{eqn-e33}. (a) The $clamped$--$ion$ contribution, and (b) the internal--strain related contribution as shown in the second term of Eq. \ref{eqn-e33}. For comparison, the data of (Yb,Al)N\cite{Jia2021} and (Yb,In)N are also provided.}
\label{e33}
\end{center}
\end{figure}

\begin{figure}[h]
\begin{center}
\includegraphics[clip, width=12cm]{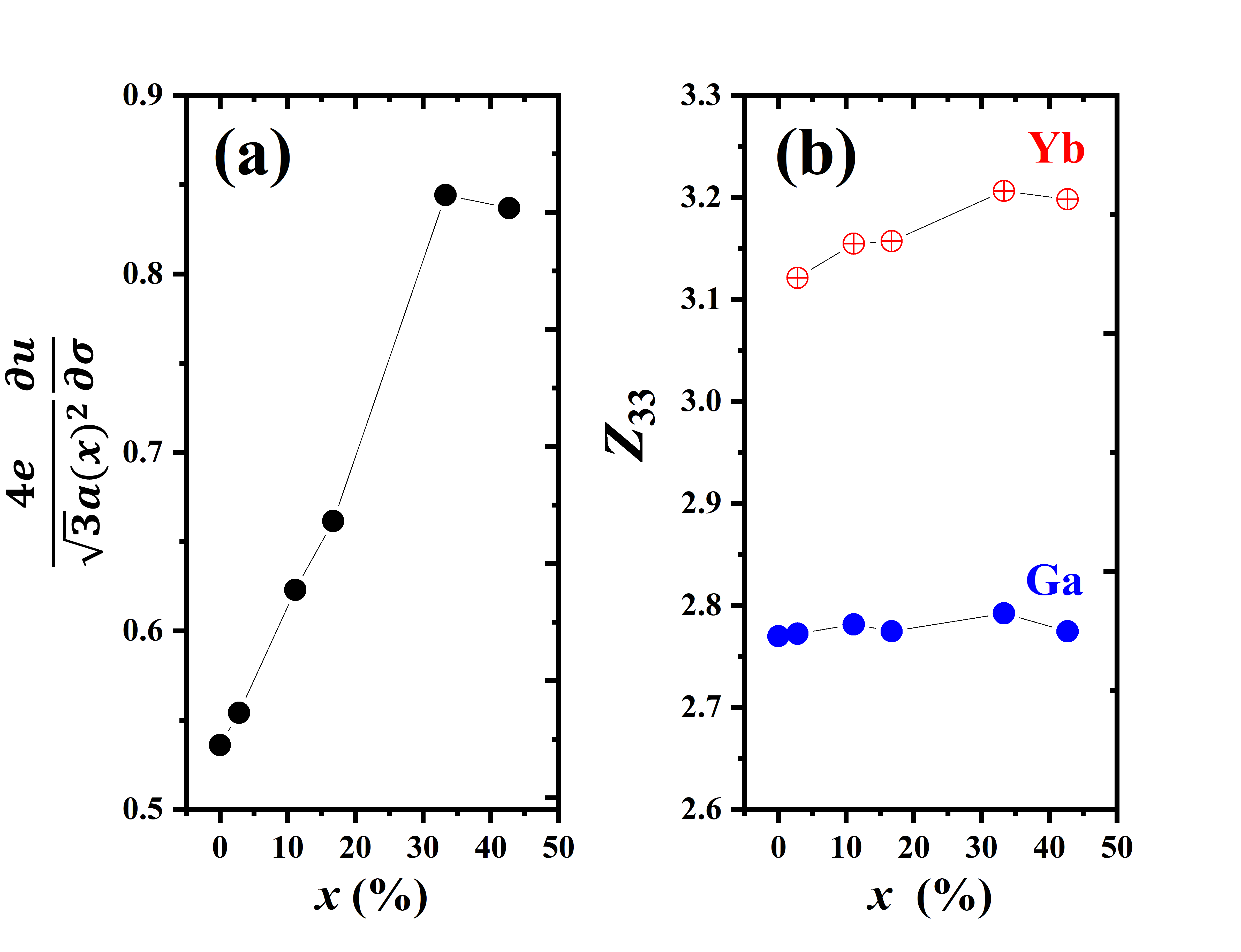}
\caption{(a) the contribution from the internal strain ($\frac{4e}{\sqrt{3}a(x)^2}\frac{\partial u}{\partial \sigma}$) to piezoelectric response, and (b) the site--resolved Born effective charges in (Yb,Ga)N.}
\label{strain}
\end{center}
\end{figure}

Fig. \ref{e33} (a) shows that the $clamped$--$ion$ term is more negative in (Yb,Ga)N than in (Yb,Al)N, and a slow increase from --1.05 to --0.89 C/m$^2$ was observed for the $clamped-ion$ term. In contrast, Fig. \ref{e33} (b) shows a fast increase in $e_{33}^{\mathrm{int}}$ with increasing the Yb concentration, where $e_{33}^{\mathrm{int}}$ represents the local structural sensitivity to macroscopic axial strain $\sigma$. To further understand the contribution of Yb substitution to $e_{33}^{\mathrm{int}}$, we calculated the  {\it internal--strain} contribution ($\frac{4e}{\sqrt{3}a(x)^2}\frac{\partial u}{\partial \sigma}$) and the site--resolved dynamic Born charge in Eq. (\ref{eqn-e33}). Fig. \ref{strain} displays that the {\it internal--strain} contribution rapidly increases with the Yb concentration, indicating that Yb substitution causes obvious local structure distortion and thus the increased piezoelectric response. On the other hand, $Z_{33}$ around the Ga site maintains at $\sim$2.80, which varies within approximately 7\% around the ionic nominal value of 3. Around the Yb sites, $Z_{33}$ shows a monotonic increase from 3.1 and 3.2. Herein, $Z_{33}<3$ may be interpreted as ``smaller ionicity" in an ionic picture, whereas $Z_{33}>3$ is due to covalency and correlation contributions.\cite{Bernardini1997} Taking all into account, the above comparisons suggest that the change in $e_{33}^{\mathrm{int}}$ in (Yb,Ga)N is mainly dominated by local structural distortion and the increased dynamic Born charge around the Yb sites caused by alloying YbN.

\subsection{Material design of piezoelectric nitride alloy}

\begin{figure}[h]
\begin{center}
\includegraphics[clip, width=12cm]{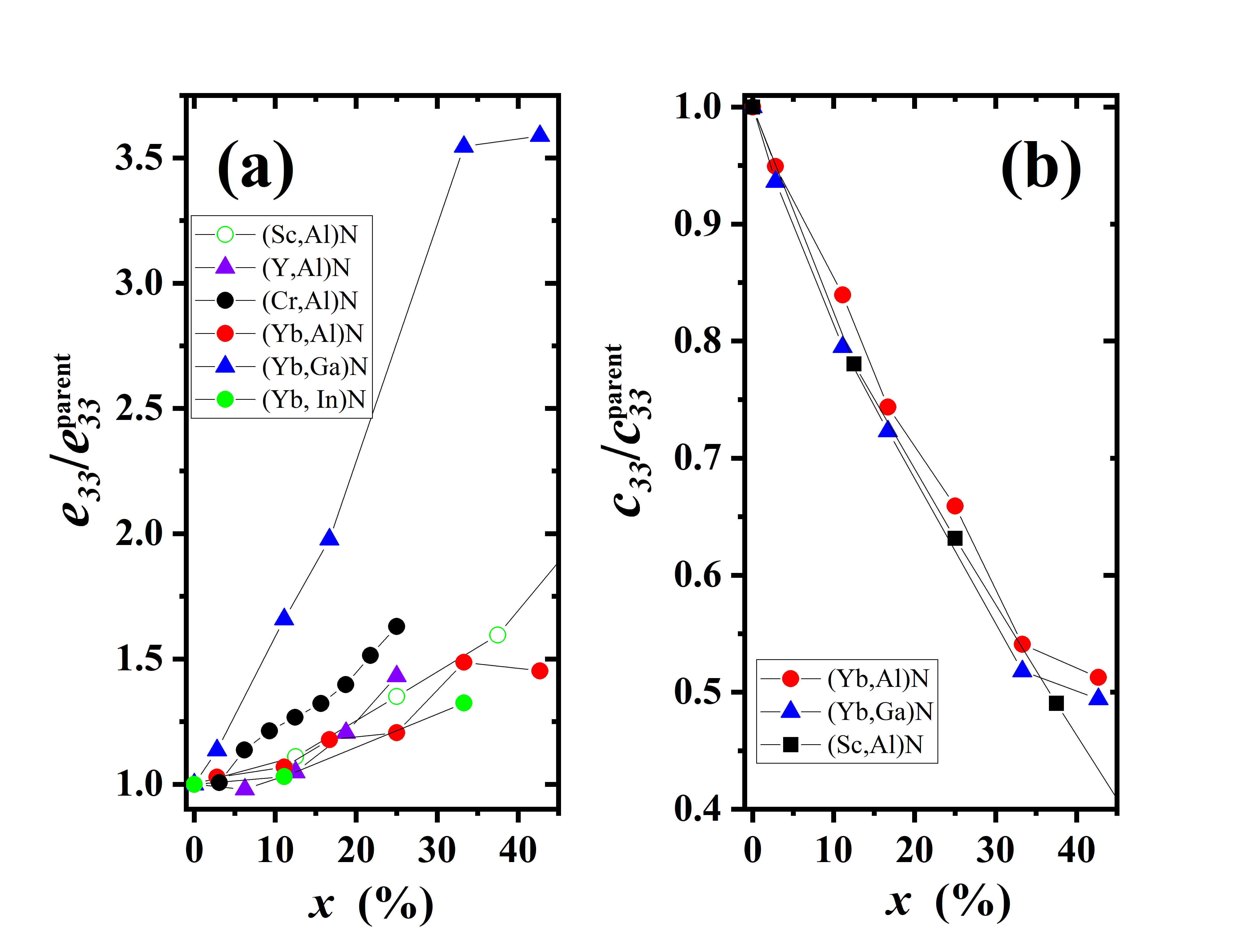}
\caption{Comparison of relative change in $e_{33}$ and $c_{33}$ to their parent nitrides with the addition of different transition metals in the $x<$45\% regime, namely $e_{33}$/$e_{33}^\mathrm{parent}$ (a) and $c_{33}$/$c_{33}^\mathrm{parent}$ (b) as a function of $x$. Here, $e_{33}^\mathrm{parent}$ and $c_{33}^\mathrm{parent}$ are elastic and piezoelectric constants of parent binary III--V nitrides, respectively. The calculated $e_{33}$ and $c_{33}$ values from other groups are also provided for comparison. (Sc,Al)N: Ref. \cite{Tasnadi2010}; (Y,Al)N and (Cr,Al)N: Ref. \cite{Manna2018}; (Yb,Al)N: Ref.\cite{Jia2021}.}
\label{comparison}
\end{center}
\end{figure}

The electromechanical coupling coefficient $k_t^2$ is approximated with the relation $k_{33}^2 \sim \frac{e_{33}^2}{\epsilon_{33} c_{33}}$ presented in Eq. \ref{k33}. For an parent III--V nitride, a large enhancement for the $k_{33}^2$ requires high $e_{33}$/$e_{33}^{\mathrm{parent}}$ and low $c_{33}$/$c_{33}^{\mathrm{parent}}$ values, simultaneously. How can we select the substitutional element for the host cation? To gain insight into identifying potential novel piezoelectric nitride materials, we collected $e_{33}$/$e_{33}^{\mathrm{parent}}$ and $c_{33}$/$c_{33}^{\mathrm{parent}}$ for different nitride alloys (see plots in Fig. \ref{comparison}).

Fig. \ref{comparison} (a) shows the $e_{33}$/$e_{33}^{\mathrm{parent}}$ of III--V nitride alloys with different substitutional elements up to 43\%. For a given parent nitride AlN, the well--studied substitutional elements such as Sc, Y, and Yb yield similar $e_{33}$/$e_{33}^{\mathrm{AlN}}$, indicating that they obey the same physical mechanism to enhance $e_{33}$, especially at $x<30\%$. This finding is unique. Since Sc has $Z_{33}<3$ in (Sc,Al)N\cite{Tasnadi2010} and Yb has $Z_{33}>3$ in (Yb,Al)N,\cite{Jia2021} Sc is considered to have a stronger ionicity than Yb in AlN. Surprisingly, such a difference in ionicity can not lead to the difference in $e_{33}$/$e_{33}^{\mathrm{AlN}}$ for (Yb,Al)N and (Sc,Al)N in Fig. \ref{comparison} (a), suggesting that the influence of the bonding state of substitutional elements on the piezoelectric response is minor. On the other hand, Fig. \ref{comparison} (b) shows that $c_{33}$/$c_{33}^{\mathrm{parent}}$ almost exhibits similar softening behavior up to 35\% for Yb-- and Sc--substituted AlN. These similar behaviors in mechanical softening in $c_{33}$ and enhancement in $e_{33}$ imply that the wurtzite AlN is an excellent material platform with a high tolerance for the impurity substitution to improve the electromechanical coupling coefficient. Notably, Cr substitution produces obviously large $e_{33}$/$e_{33}^{\mathrm{parent}}$ in AlN. Manna et al.\cite{Manna2018} explained that it can be attributed to the increase of the internal parameter $u$ of the wurtzite structure upon substitution of Al with larger Cr ions partly because of the larger ionic radius of Cr than that of Al. However, this seems irrational because other substitutional elements have larger ionic size with coordination number of 6 (Sc$^{3+}$: 74.5 pm, Yb$^{3+}$: 86.8 pm, and Y$^{3+}$: 90.0 pm)\cite{Shannon1976}.

When the host nitride changes from AlN to GaN, Yb--substitution induces an obvious relative change in $e_{33}$/$e_{33}^{\mathrm{parent}}$, as shown in Fig. \ref{comparison} (a). The following question is why alloying YbN into GaN shows an obvious change in $e_{33}$/$e_{33}^{\mathrm{parent}}$ compared with other AlN--based alloys as shown in Fig. \ref{comparison} (a). In this study, $e_{33}$/$e_{33}^{\mathrm{parent}}$ is defined as the ratio of $e_{33}$ value of nitride alloy to $e_{33}^{\mathrm{parent}}$ of its parent nitride, where $e_{33}$ can be further decomposed into $e_{33}^{\mathrm{clamped}}$ and $e_{33}^{\mathrm{int}}$ as follows: 
\begin{equation}
\frac{e_{33}}{e_{33}^{\mathrm{parent}}}=\frac{e_{33}^{\mathrm{clamped}}+e_{33}^{\mathrm{int}}}{e_{33}^{\mathrm{parent}}}.  
\label{comp}
\end{equation}

\begin{figure}[h]
\begin{center}
\includegraphics[clip, width=12cm]{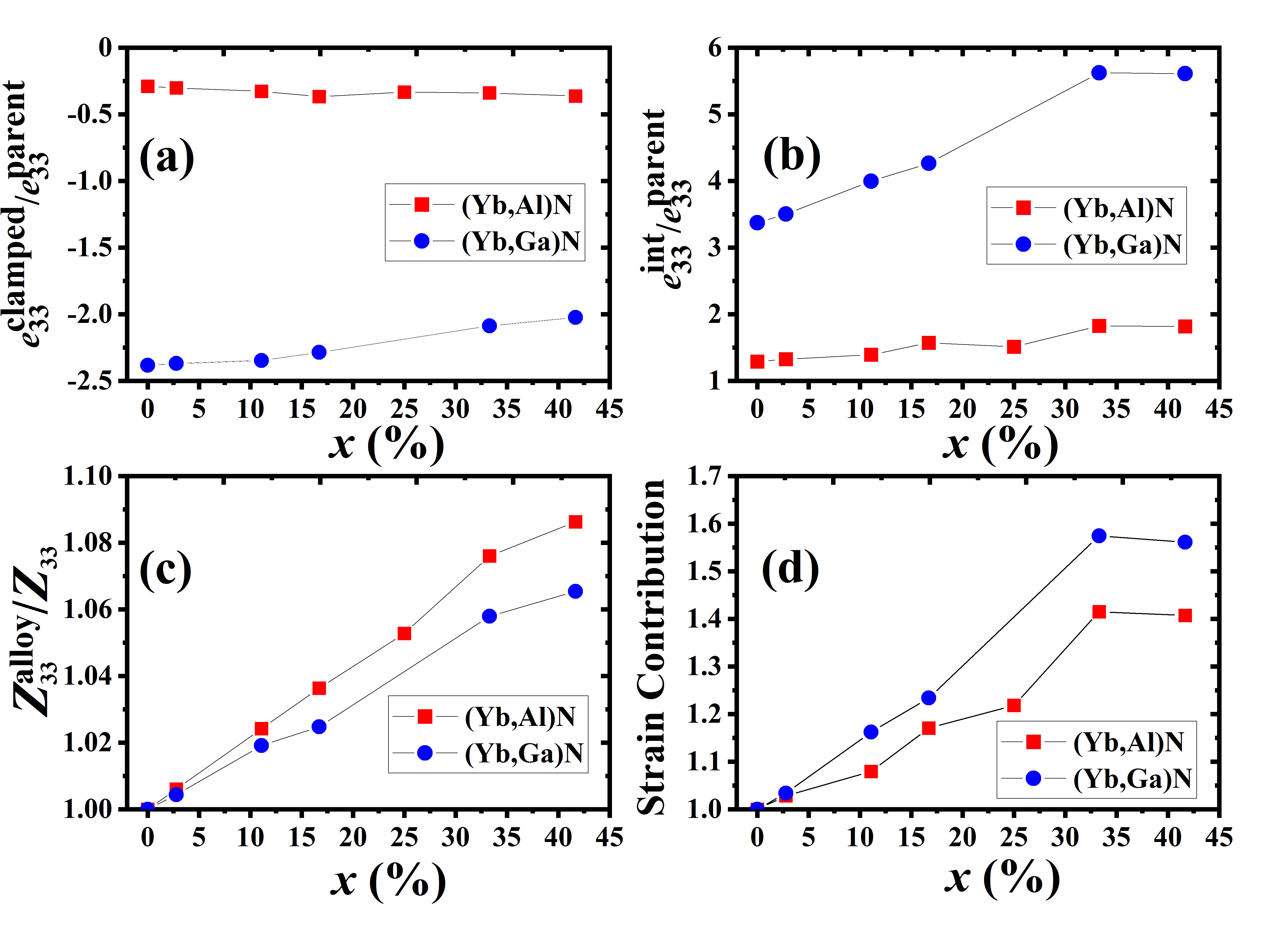}
\caption{The decomposed contributions for $e_{33}$/$e_{33}^{\mathrm{clamped}}$ as introduced by Eq. (7): (a) represents the contribution of the {\it clamped-ion} term ($e_{33}^{\mathrm{clamped}}$/$e_{33}^{\mathrm{parent}}$); (b) indicates the {\it internal--strain} ($e_{33}^{\mathrm{int}}$/$e_{33}^{\mathrm{parent}}$. (c) is the relative changes of dynamical Born change to their nitride parents, namely ($Z_{33}^{\mathrm{alloy}}$/$Z_{33}^{\mathrm{parent}}$); (d) shows the relative contribution of {\it internal--strain} ($\frac{4e}{\sqrt{3}a(x)^2}\frac{\partial u}{\partial \sigma}$) to their nitride parents, in both (Yb,Al)N and (Yb,Ga)N.}
\label{Fig9}
\end{center}
\end{figure}

Figure \ref{Fig9} (a) and (b) show the decomposed contributions from $e_{33}^{\mathrm{clamped}}$ and $e_{33}^{\mathrm{int}}$ to $e_{33}$/$e_{33}^{\mathrm{parent}}$, where $e_{33}^{\mathrm{AlN}}$ and $e_{33}^{\mathrm{GaN}}$ of the parent nitrides AlN and GaN are set to 1.46 and 0.44 C/m$^2$ from our calculations, respectively.\cite{Jia2021} Fig. \ref{Fig9}  (a) shows that $e_{33}^{\mathrm{clamped}}$/$e_{33}^{\mathrm{parent}}$ has less change with the Yb concentration in both (Yb,Ga)N and (Yb,Al)N. In contrast, Fig. \ref{Fig9} (b) shows that $e_{33}^{\mathrm{int}}$/$e_{33}^{\mathrm{parent}}$ in (Yb,Ga)N is obviously larger than that in (Yb,Al)N, which is considered as a main factor to cause a large change in $e_{33}$/$e_{33}^{\mathrm{parent}}$ in (Yb,Ga)N. On the other hand,  $e_{33}^{\mathrm{int}}$ in both (Yb,Ga)N and (Yb,Al)N have a similar tendency with increasing the YbN concentration, and their values also become close at high Yb concentrations as shown in Fig. \ref{e33} (b). Eventually, considering a  close $e_{33}^{\mathrm{int}}$ in (Yb,Ga)N and (Yb,Al)N,  an obviously large $e_{33}$/$e_{33}^{\mathrm{parent}}$ in GaN-based nitride alloys is mainly attributed to the small $e_{33}^{\mathrm{GaN}}$ (0.44 C/m$^2$) of the parent GaN.

On the other hand, $c_{33}$/$c_{33}^{\mathrm{parent}}$ shows a similar tendancy with increasing Yb concentration up to $x\sim$40\% in (Yb,Al)N and (Yb,Ga)N, even including (Sc,Al)N as shown in Fig. \ref{comparison} (b), implying that their mechanical softening obeys the same physical mechanism. For (Yb,Al)N and (Yb,Ga)N, the mechanical softening is partially attributed to the increased Yb--Yb pair interaction due to the formation of --Yb--N--Yb-- chain along the $c$--axis, as revealed in (Yb,Al)N.\cite{Jia2021} Such softening mechanism is also expected in (Sc,Al)N. From Eq. (5), it is known that the mechanical softening has a key influence on $k_{33}^2$. However, the $k_{33}^2$ value in (Yb,Ga)N (see Table \ref{cij}) is obviously smaller than that in (Yb,Al)N at the same concentration,\cite{Jia2021} \textit{e.g.}, 2.62\% in (Yb,Ga)N versus 11.52\% in (Yb,Al)N at $x$=17\%. This small $k_{33}^2$ in (Yb,Ga)N is attributed to its small $e_{33}$. Further calculations show that it is mainly caused by an obviously small \textit{clamped--ion} term in GaN, \textit{e.g.}, --1.03 C/m$^2$ in (Yb,Ga)N versus --0.48 C/m$^2$ in (Yb,Al)N at $x$=11\%, as shown in Table \ref{u-value}.

\begin{table}[ht]
\caption{The $clamped$--$ion$ ($e_{33}^{\mathrm{clamped}}$ in C/m$^2$) and internal strain ($e_{33}^{\mathrm{int}}$ in C/m$^2$) contributions versus the wurtzite internal parameter of III--V nitride alloys. Here, $u$ is the average wurtzite internal parameter, and $u_{\mathrm{M}}$ (M=Ga, In, Al, Yb) represents the site--resolved $u$ value.}
\begin{adjustbox}{width=1\textwidth}
\begin{tabular}{l|ccccc|ccccc|ccccc}
\hline
& \multicolumn{5}{c}{GaN}    &  \multicolumn{5}{c}{InN}   &    \multicolumn{5}{c}{AlN}\\
\cline{2-6}\cline{7-11}\cline{12-16}
       & $e_{33}^{\mathrm{clamped}}$ & $e_{33}^{\mathrm{int}}$  & $u_{\mathrm{Ga}}$    & $u_{\mathrm{Yb}}$ & $u$ & $e_{33}^{\mathrm{clamped}}$ & $e_{33}^{\mathrm{int}}$ & $u_{\mathrm{In}}$  & $u_{\mathrm{Yb}}$ & $u$ & $e_{33}^{\mathrm{clamped}}$ & $e_{33}^{\mathrm{int}}$ &$u_{\mathrm{Al}}$  & $u_{\mathrm{Yb}}$ & $u$ \\
\hline
$x$=0.00   & -1.05 & 1.48 & 0.377  & --     &0.377   & -0.95   & 1.91      & 0.379   &  --      &0.379    & -0.42   & 1.88   &0.382   &  --    & 0.382 \\
$x$=0.03   & -1.04 & 1.54 & 0.376  & 0.413 & 0.377  & -0.96   & 1.95      & 0.379   & 0.380   & 0.380   & -0.43   & 1.94   & 0.381  & 0.432 & 0.382 \\
$x$=0.11   & -1.03 & 1.76 & 0.375  & 0.411 &0.379   & -1.03   & 2.02      & 0.381   & 0.380   &0.381   & -0.48   & 2.03   & 0.377  & 0.445  & 0.384 \\
$x$=0.33   & -0.91 & 2.47 & 0.375  & 0.415 &0.388   & -0.85   & 2.13      & 0.386   & 0.385   &0.386   & -0.50   & 2.67   & 0.377  & 0.431  & 0.395 \\
\hline
\end{tabular} 
\end{adjustbox}
\label{u-value}
\end{table}

The \textit{clamped--ion} term represents the effect of the external strain on the electronic structure, i.e., the ionic coordinates follow the strain--induced deformation of the lattice vectors.\cite{Wang2021} Table \ref{u-value} shows an increase in the \textit{clamped--ion} term together with an increased $u$ at the same $x$, when the structure goes from GaN to InN and then to AlN alloys. In principle, the internal parameter $u$ is defined as the length of the bond parallel to the $c$--axis (anion--cation bond length) divided by the lattice constant $c$. Thus, $u$ and $c$ are not independent, and the \textit{clamped--ion} term in Eq. (\ref{eqn-e33}) possibly depends on the $u$ value. For example, at $x$=0.00, $u$ of GaN is the smallest in three nitrides, and corresponds to the smallest \textit{clamped--ion} term. Moreover, $u$ also decribes the asymmetry of atomic arrangement along the $c$--axis in the wurtzite, \textit{e.g.}, $\cdots$Al--N$\cdots$Al--N$\cdots$, where the symbol ($\cdots$) represent the nonbonding state. For an ideal wurtzite structure, $u$ is 0.375. The increasing nonideality for $u$ from (Yb,Ga)N to (Yb,In)N to (Yb,Al)N at the same $x$ is considered as intrinsic structural properties of these alloys. After the axial strain is applied parallel to the $c$--axis, a large structural nonideality is considered to generate a large electronic response due to the asymmetry of the bonding and nonbonding lengths. This also explains why using GaN as the parent nitride is difficult to improve its $k_t^2$ to a large value compared with AlN-based nitride alloys. 

It is worthy to discuss the effect of size mismatch between the substitutional and host cations on the piezoelectric response based on our calculation results. As revealed by our mixing enthalpy calculations in Fig. \ref{Entalpy}, YbN is easier to alloy into GaN than AlN, which is attributed to the size difference between the substitutional and host cations. This size effect is also observed in the reported AlN--based nitride alloys.\cite{Manna2018} With the substitutional element changing from Y to Sc to Cr (Y$^{3+}$: 90.0 pm, Sc$^{3+}$: 74.5 pm, and Cr$^{3+}$: 62.0 pm)\cite{Shannon1976}, the maximal mixing enthalpy for wurtzite structure gradually increases,\cite{Manna2018} confirming that the substitutional element with a close ionic size to the host cation is easier to substitute into the host cation site. The following question is whether such a size effect affect the piezoelectric response or not. Fig. \ref{e33} (b) shows that $e_{33}^{\mathrm{int}}$ increases with the Yb concentration for both (Yb,Ga)N and (Yb,Al)N. To understand the size effect on the increased piezoelectric response, we decomposed $e_{33}^{\mathrm{int}}$ into the {\it internal--strain} contribution ($\frac{4e}{\sqrt{3}a(x)^2}\frac{\partial u}{\partial \sigma}$) and the dynamic Born charge as given in Eq. \ref{eqn-e33}. Fig. \ref{Fig9} (c) shows that the change of dynamical Born charge of (Yb,Al)N relative to its parent AlN is larger than that in (Yb,Ga)N. On the contrary, Fig. \ref{Fig9} (d) indicates that the change of the {\it internal--strain} contribution relative to GaN in (Yb,Ga)N is larger than that in (Yb,Al)N. Considering together with a larger increase in $e_{33}$/$e_{33}^{\mathrm{parent}}$ in (Yb,Ga)N in Fig. \ref{comparison} (a), it is concluded that Yb substitution in (Yb,Ga)N can induce a larger {\it internal--strain} contribution than that in (Yb,Al)N, which is attributed to the size difference between the substitutional and host cations. Note that a large {\it internal--strain} contribution in (Yb,Ga)N also leads to the enhanced $e_{33}$/$e_{33}^{\mathrm{parent}}$ in (Yb,Ga)N.

Based on various theoretical calculation results in Fig. \ref{comparison} and the above discussions, we proposed a simple guideline to select alloying components in a search for a massive increase in electromechanical coupling. First, the core guideline is to select a host wurtzite material with the \textit{clamped--ion} term as close to zero as possible. The parent nitride with an excessively negative \textit{clamped--ion} term leads to difficulty to improve the absolute magnitude of $k_t^2$, as revealed in GaN--based nitride alloys. Our calculation results show that a smaller internal parameter $u$ is associated with a more negative \textit{clamped--ion} term for III--V nitrides, where tailoring $u$ is possible by adjusting the lattice parameter $c$ or the cation--anion bond length in the $c$ axis based on the impurity substitution as shown in Table \ref{u-value}. Second, selecting the substitutional element with an optimized ionic size is critical. A suitable size mismatch can give rise to an enhanced internal strain, and thus an enhanced piezoelectric response. 

The most straightforward implication of our work is to identify potential novel piezoelectric nitride materials. However, identifying nitride alloys with high $e_{33}$/$e_{33}^{\mathrm{parent}}$ and low $c_{33}$/$c_{33}^{\mathrm{parent}}$ values are two necessary conditions, rather than sufficient conditions for piezoelectric applications. Good piezoelectric materials would require satisfying other constraints such as the phase stability and wide bandgap. Follow--up phase stability studies are required on the candidates with high $e_{33}$/$e_{33}^{\mathrm{parent}}$ and low $c_{33}$/$c_{33}^{\mathrm{parent}}$, which will help determine the fabrication process. Besides, for material discovery, the bandgap of piezoelectric materials should be considered. For example, the parent nitride InN has a bandgap of 0.8 eV, and such a narrow bandgap limits its use in piezoelectric energy harvesters and related devices because valence electrons are easy to pass through a narrow bandgap by thermal excitation. Furthermore, the isovalent substitution needs to avoid orbital overlap with host elements and does not influence the conduction band character of the parent wurtzite structure. For aliovalent substitution, the generation of charged carriers must be avoided. Finally, we should stress that we assumed a single substitutional cation to be the only dopant. When an additional dopant is present, further enhancement in $e_{33}$/$e_{33}^{\mathrm{parent}}$ beyond the solubility of a single dopant is expected. Alloying with two or more species into III--V nitrides may open a valuable route for simultaneously engineering the piezoelectric and elastic properties of a parent wurtzite structure, \textit{e.g.}, (Y, B, Al)N calculated by Manna et al.\cite{Manna2017} 

\section{Summary}

The enhancement of electromechanical coupling due to Yb substitution in III--V nitrides was investigated. The Yb substitution induces an increase in piezoelectric coefficients accompanied by a substantial decrease in elastic moduli and thus enhances the electromechanical coupling coefficient compared with the parent GaN or AlN nitride. Our calculations revealed that the enhancement of piezoelectric response induced by the substitutional elements such as Sc, Y, and Yb almost obeys the same physical mechanism, viz, to enhance the $internal$ $strain$. The strongest enhancement in piezoelectric response relative to the parent nitride is observed in (Yb,Ga)N, which is mainly attributed to a fairly small $e_{33}$ of parent GaN. The substitutional element with a closer ionic size to the host cation is easier to substitute into the host cation site, and produces a larger internal strain to partly contribute to the enhancement in piezoelectric response. Likewise, the Yb--substitution induced decrease in $c_{33}$ is also considered to follow the same softening mechanism in both GaN and AlN. Our work suggests that the matching of ionic size between substitutional and host cations can serve as a guideline of material design for determining alloying components in a search with the aim of enhanced electromechanical coupling, and provide new insights into material design strategies for III--V nitride piezoelectric materials, such as multiple element substitutions.

\section{Acknowledgement} 
J. Jia acknowledges the funding from JSPS KAKENHI Grant-in-Aid for Scientific Research (C) (Grant No. 20K05368), and from Waseda University Grant for Special Research Projects (Project number: 2020C--316).  T. Yanagitani thanks the support from JST CREST (Grant No. JPMJCR20Q1), Japan.


\begin{thebibliography}{9}
\bibitem{Jia2021} Jia J.; Yanagitani T.; Origin of Enhanced Electromechanical Coupling in (Yb,Al)N Nitride Alloys. {\it Phys. Rev. Applied} {\bf 2018}, 16, 044009.

\bibitem{Yanagitani2014} Yanagitani T.; Suzuki M.; Enhanced Piezoelectricity in YbGaN Films Near Phase Boundary. {\it Appl. Phys. Lett.} {\bf 2014}, 104, 082911.

\bibitem{Akiyama2009} Akiyama M.; Kamohara T.; Kano K.; Teshigahara A.; Takeuchi Y.; Kawahara N.; Enhancement of Piezoelectric Response in Scandium Aluminum Nitride Alloy Thin Films Prepared by Dual Reactive Cosputtering. {\it Adv. Mater.} {\bf 2009}, 593, 21.

\bibitem{Akiyama2009-2} Akiyama M.; Kano K.; Teshigahara A.; Influence of Growth Temperature and Scandium Concentration on Piezoelectric Response of Scandium Aluminum Nitride Alloy Thin Films. {\it Appl. Phys. Lett.} {\bf 2009}, 162107, 95.

\bibitem{Tholander2013} Tholander C.; Abrikosov I. A.; Hultman L.; Tasn\'{a}di F.; Volume Matching Condition to Establish the Enhanced Piezoelectricity in Ternary (Sc,Y)$_{0.5}$(Al,Ga,In)$_{0.5}$N alloys. {\it Phys. Rev. B} {\bf 2013}, 094107, 87.

\bibitem{Tholander2015} Tholander C.; Tasn\'{a}di F.; Abrikosov I. A.; Hultman L.; Birch J.; Alling B.; Large Piezoelectric Response of Quarternary Wurtzite Nitride Alloys and Its Physical Origin From First Principles. {\it Phys. Rev. B} {\bf 2015}, 174119, 92.

\bibitem{Jia2017} Isosaki Y.; Yamashita Y.; Yagi T.; Jia J.; Taketoshi N.; Nakamura S.; Shigesato Y.; Structure and Thermophysical Properties of GaN Films Deposited by Reactive Sputtering Using a Metal Ga Target. {\it J. Vac. Sci. Technol. A} {\bf 2017}, 041507, 35.

\bibitem{Mirko2016} Weidner M.; Jia J.; Shigesato Y.; Klein A.; Comparative Study of Sputter-deposited SnO$_2$ Films Doped with Antimony or Tantalum. {\it Phys. Status Solidi B} {\bf 2016}, 923, 253.

\bibitem{Jia2014} Jia J.; Yoshimura A.; Kagoya Y.; Oka N.; Shigesato Y.; Transparent Conductive Al and Ga Doped ZnO Films Deposited Using Off--axis Sputtering. {\it Thin Solid Films} {\bf 2014}, 69, 559.

\bibitem{Trumbore1960} Trumbore F. A.; Solid Solubilities of Impurity Elements in Germanium and Silicon. {\it Bell Syst. Tech. J.} {\bf 1960}, 205, 39.

\bibitem{Shannon1976} Shannon R. D.; Revised Effective Ionic Radii and Systematic Studies of Interatomic Distances in Halides and Chalcogenides. {\it Acta Crystallographica} {\bf 1976}, 751, 32.

\bibitem{Kresse1996} G. Kresse and J. Furthm\"{u}ller, Comput. Mater. Sci. {\bf 6}, 15 (1996).

\bibitem{PBE} J. P. Perdew, K. Burke, and M. Ernzerhof, Phys. Rev. Lett.  {\bf 77},  3865 (1996).

\bibitem{Zunger1990} A. Zunger, S.-H. Wei, L. G. Ferreira, and J. E. Bernard, Phys. Rev. Lett. {\bf 65}, 353 (1990).


\bibitem{Edgar1994} Edgar J. H., ed., Properties of Group-III Nitrides and EMIS Data Reivews (IEE, London, 1994).

\bibitem{Nakamura2012} Nakamura N.; Ogi H.; Hirano M.; Elastic, Anelastic, and Piezoelectric Coefficients of GaN. {\it J. Appl. Phys.} {\bf 2012}, 013509, 111.

\bibitem{Alling2007} Alling B.; Ruban A. V.; Karimi A.; Peil O. E.; Simak S. I.; Hultman L.; Abrikosov I. A.; Mixing and Decomposition Thermodynamics of $c$--Ti$_{1-x}$Al$_x$N from First--Principles Calculations. {\it Phys. Rev. B} {\bf 2007}, 045123, 75.

\bibitem{Talley2018} Talley K. R.; Millican S. L.; Mangum J.; Siol S.; Musgrave C. B.; Gorman G.; Holder A. M.; Zakutayev A.; Brennecka G. L.; Implications of heterostructural alloying for enhanced piezoelectric performance of (Al,Sc)N. {\it Phys. Rev. Materials} {\bf 2018}, 063802, 2.

\bibitem{Zhang2013} Zhang S.; Fu W. Y.; Holec D.; Humphreys C. J.; Moram M. A.; Elastic Constants and Critical Thicknesses of ScGaN and ScAlN. {\it J. Appl. Phys.} {\bf 2013}, 243516, 114.

\bibitem{Zukauskaite2012} Zukauskaite A.; Wingqvist G.; Palisaitis J.; Jensen J.; Persson P.; Matloub R.; Muralt P.; Yunseok K.; Birch J.; Hultman L.; Microstructure and Dielectric Properties of Piezoelectric Magnetron Sputtered $w$--Sc$_x$Al$_{1-x}$N Thin Films {\it J. Appl. Phys.} {\bf 2012}, 093527, 111.

\bibitem{Hoglund2009} H\"{o}glund C.; Bare\~{n}o J.; Birch J.; Alling B.; Czig\'{a}ny Z.; Hultman L.; Cubic Sc$_{1-x}$Al$_x$N solid solution thin films deposited by reactive magnetron sputter epitaxy onto ScN(111) {\it J. Appl. Phys.} {\bf 2009}, 113517, 105.

\bibitem{Feneberg2007} Feneberg M.; Thonke K.; Polarization Fields of III--Nitrides Grown in Different Crystal Orientations. {\it J. Phys. D: Appl. Phys.} {\bf 2007}, 403201, 19.

\bibitem{Bernardini1997} Bernardini F.; Fiorentini V.; Vanderbilt D.; Spontaneous Polarization and Piezoelectric Constants of III-V Nitrides. {\it Phys. Rev. B} {\bf 1997}, R10024, 56.

\bibitem{Wang2021} Wang H.; Adamski N.; Mu S.; Van de Walle C. G.; Piezoelectric effect and polarization switching in Al$_{1-x}$Sc$_x$N. {\it J. Appl. Phys.} {\bf 2021}, 104101, 130.

\bibitem{Tasnadi2010} Tasn\'{a}di F.; Alling B.; Hoglund C.; Wingqvist G.; Birch J.; Hultman L.; Abrikosov I. A.; Origin of the Anomalous Piezoelectric Response in Wurtzite Sc$_x$Al$_{1-x}$N Alloys. {\it Phys. Rev. Lett.} {\bf 2010}, 137601, 104.

\bibitem{Manna2018} Manna S.; Talley K. R.; Gorai P.; Mangum J.; Zakutayev A.; Brennecka G. L.; Stevanovi\'{c} V.; Ciobanu C. V.; Enhanced Piezoelectric Response of AlN via CrN Alloying. {\it Phys. Rev. Applied} {\bf 2018}, 034026, 9.

\bibitem{Manna2017} Manna S.; Brennecka G. L.; Stevanovi\'{c} V.; Ciobanu C. V.; Tuning the Piezoelectric and Mechanical Properties of the AlN System via Alloying with YN and BN. {\it J. Appl. Phys.} {\bf 2017}, 105101, 122.

\end{thebibliography}
\end{document}